\documentclass{article}

\usepackage[main, final]{iaseai26}

\usepackage[utf8]{inputenc} 
\usepackage[T1]{fontenc}    
\usepackage{hyperref}       
\usepackage{url}            
\usepackage{booktabs}       
\usepackage{amsfonts}       
\usepackage{nicefrac}       
\usepackage{microtype}      
\usepackage{xcolor}         
\usepackage{float}

\usepackage[dvipsnames]{xcolor}
\usepackage{tikz}
\usetikzlibrary{positioning, arrows.meta, calc, fit}

\definecolor{riskbg}{RGB}{220,220,220}
\definecolor{prevbg}{RGB}{210,230,250}
\definecolor{mitibg}{RGB}{255,230,200}
\definecolor{gapred}{RGB}{200,60,60}

\definecolor{scenariobg}{RGB}{230,240,250}
\definecolor{filingbg}{RGB}{250,240,220}
\definecolor{srrbg}{RGB}{220,245,220}
\definecolor{crisisbg}{RGB}{245,230,245}
\definecolor{arrowcol}{RGB}{80,80,80}
\definecolor{feedbackcol}{RGB}{150,100,100}

\title{The Coordination Gap in Frontier AI Safety Policies}

\author{%
  Isaak Mengesha\\
  Informatics Institute\\
  University of Amsterdam\\
  Amsterdam, 1098XH\\
  \texttt{i.a.mengesha@uva.nl} \\
}

\begin{document}

\maketitle

\begin{abstract}
Frontier AI Safety Policies concentrate on prevention: capability evaluations, deployment gates, usage constraints, while neglecting institutional capacity to coordinate responses when prevention fails. We argue this coordination gap is structural: investments in ecosystem robustness yield diffuse benefits but concentrated costs, generating systematic underinvestment. Drawing on risk regimes in nuclear safety, pandemic preparedness, and critical infrastructure, we propose that similar mechanisms (precommitment, shared protocols, standing coordination venues) could be adapted to frontier AI governance. Closing the gap requires cross-actor "note-exchange" of ex ante if-then response logic, exposing not only triggers but the decision processes that convert signals into actions. Without such architecture, institutions cannot learn from failures at the pace of relevance.
\end{abstract}

\section{Introduction}

The central challenge of AI governance is to harness the economic and social benefits of widespread deployment while preventing the potential harms that such systems may generate. Yet the societal impacts of advanced AI systems remain difficult to anticipate \cite{anderljung2023frontier}, creating fundamental tensions in policy design. Regulators seek to hold developers accountable without driving innovation offshore, and to impose sufficient safety requirements without stifling the very technological progress they aim to govern. This is a delicate balancing act, and most research has naturally focused on optimizing preventative measures for development and deployment \cite{weidinger2023sociotechnical}.

Given the scale of potential harms, societies are likely underinvesting in AI safety, and additional preventive effort plausibly yields positive marginal returns \cite{jones2025much}. Yet even prevention with extremely high reliability, e.g. on the order of 99\%, is unlikely to guarantee full containment in complex socio-technical systems \cite{perrow2011normal}. Recognizing this, many researchers anticipate that harmful incidents or near-misses (“warning shots” or “focusing events”) will be necessary to shift political feasibility toward more stringent and costly interventions \cite{birkland1997after}. In effect, this requires taking seriously the possibility of failures that generate large or even catastrophic societal harm. As AI capabilities and deployment accelerate, such failures become more likely.

Importantly, this does not weaken the case for prevention: preventive investment continues to dominate expected-value calculations for existential risk. The question is instead whether accounting for inevitable failures changes where additional effort has the highest marginal payoff. We argue that it does. When institutions lack capacity for shared learning, coordinated response, and adaptive adjustment, governance is forced to rely almost exclusively and increasingly on preventive control, even when prevention cannot eliminate residual risk. Under these conditions, the absence of resilience is what makes prevention the only available, yet ultimately insufficient, strategy.

Broader frameworks for societal adaptability have been proposed \cite{bernardi2024societal}, yet they leave open concrete implementation pathways. Meanwhile, a large body of work has identified specific harms: cognitive overload at the societal level \cite{lahlou2025mitigating} and shifts in the offense-defense balance in cybersecurity \cite{garfinkel2021does}, among others. More broadly, the literature identifies a ``context gap'' between technical prevention protocols and holistic societal impact assessments \cite{weidinger2023sociotechnical}. What matters for frontier-AI governance is whether systems and institutions can maintain acceptable performance under unexpected conditions and continue functioning when shocks occur. Given the costs of such shocks, the goal is to minimize their occurrence, maximize learning from them, and where possible learn even in their absence. Current policies address only the first \cite{capano2017resilience}.

Many proposals contribute frameworks for monitoring and reflection around such events \cite{gandhi2025societal, solaiman2023evaluating, UKAISI2025}. Monitoring is indeed a necessary condition for effective mitigation. Yet the underlying challenge is a collective action problem: benefits are diffuse while costs are concentrated, creating incentives to free ride and a high risk of coordination failure. These dilemmas are well studied, and established strategies exist to address them; one such mechanism is precommitment \cite{ostrom1990governing}.

The relevant actors span government regulators, critical infrastructure operators, private labs and cloud providers, and civil society. No single actor possesses authority or control across all domains, yet optimal responses require coordination among them. During high-pressure crisis events, this coordination capacity becomes critical, a point recognized by calls for interagency task forces, public-private partnerships, and treaty-like mechanisms. From other high-risk fields, we can extract best practices for such coordination efforts \cite{stetler2025reinsuring}.

While multiple actors must account for robustness to improve coordination, labs as primary developers and operators carry particular responsibility. Labs are already producing Frontier AI Safety Policies (FASPs), yet these policies implicitly assume that preventative measures will succeed. Taking seriously the possibility that they will not—and planning accordingly—requires integrating robustness standards into these policies, making them the natural anchor point for multi-actor coordination.\\

\section{The 'robustness gap' in the broader Ecosystem}

Robustness in frontier-AI governance concerns whether policies and institutions continue to function when underlying assumptions fail, signals are ambiguous, or external conditions shift in ways developers did not anticipate \cite{mens2011meaning}. The problem is that current approaches operate under the assumption of control and fail to take failure seriously.

\subsection{Consequences of ignoring robustness}

Traditional ``predict-then-act'' approaches assume accurate forecasts. Under conditions of deep uncertainty, this produces brittle policies that lead to one of two failure modes: gridlock, where actors cannot agree on underlying assumptions, or overconfidence, where plans fail when futures diverge from expectations \cite{randRDMtool,lempert2013making}. This brittleness is particularly consequential when actors operate under differing assumptions about baseline risk and responsibility \cite{fli2025aisafetyindex, field2025experts}. While actors may not be able to agree on underlying probability distributions---and therefore on adequate degrees of preventative measures---they can nevertheless disclose their post-hoc responses in light of impending harm or novel evidence of potential harm \cite{NRC2009Informing,Haasnoot2013}.

The myopic focus on monitoring compliance of individual developers neglects socio-technical feedback loops. System-level harms such as cyber-risks or disinformation cascades cannot be mitigated by individual actors alone \cite{moran2023critical}. When policies meet internal compliance standards yet still fail under real-world shocks, trust in governance architectures declines. Risk management in low-trust environments poses major challenges, as documented in other high-risk contexts \cite{OECD2017Trust, IRGC2017, NAIIC2012}. 

\begin{figure}[htbp]
\centering

\begin{tikzpicture}[
    node distance=0.4cm and 0.6cm,
    cell/.style={
        rectangle,
        rounded corners=3pt,
        minimum width=2.4cm,
        minimum height=0.9cm,
        align=center,
        font=\footnotesize,
        draw=black!50,
        thick
    },
    rowlabel/.style={
        font=\footnotesize\bfseries,
        minimum width=1.8cm,
        align=right
    },
    causal/.style={
        -{Stealth[length=5pt]},
        thick,
        black!70
    }
]

\node[rowlabel] (labelRisk) at (0, 2.4) {Risk};
\node[rowlabel] (labelPrev) at (0, 1.2) {Prevention};
\node[rowlabel] (labelMiti) at (0, 0) {Mitigation};

\def\colA{2.2}
\def\colB{5.0}
\def\colC{7.8}
\def\colD{10.6}

\node[cell, fill=riskbg] (R1) at (\colA, 2.4) {Model\\capabilities};
\node[cell, fill=riskbg] (R2) at (\colB, 2.4) {Deployment\\scale};
\node[cell, fill=riskbg] (R3) at (\colC, 2.4) {Systemic\\exposure};
\node[cell, fill=riskbg] (R4) at (\colD, 2.4) {Societal\\harm};

\node[cell, fill=prevbg] (P1) at (\colA, 1.2) {Alignment,\\testing};
\node[cell, fill=prevbg] (P2) at (\colB, 1.2) {Usage\\policies};
\node[cell, fill=prevbg] (P3) at (\colC, 1.2) {Compliance\\audits};
\node[cell, fill=prevbg] (P4) at (\colD, 1.2) {Regulation,\\norms};

\node[cell, fill=mitibg] (M1) at (\colA, 0) {Rollback,\\patches};
\node[cell, fill=mitibg] (M2) at (\colB, 0) {Incident\\response};
\node[cell, fill=mitibg] (M3) at (\colC, 0) {Cross-sector\\coordination};
\node[cell, fill=mitibg] (M4) at (\colD, 0) {Crisis\\management};

\draw[causal] (R1.east) -- (R2.west);
\draw[causal] (R2.east) -- (R3.west);
\draw[causal] (R3.east) -- (R4.west);

\node[font=\footnotesize\bfseries, align=center] at (\colA, 3.3) {Developer/\\Lab};
\node[font=\footnotesize\bfseries, align=center] at (\colB, 3.3) {Deployer/\\Platform};
\node[font=\footnotesize\bfseries, align=center] at (\colC, 3.3) {Integrators/\\Institutions};
\node[font=\footnotesize\bfseries, align=center] at (\colD, 3.3) {End Users/\\Society};

\node[
    draw=gapred,
    dashed,
    very thick,
    rounded corners=5pt,
    fit=(M2)(M3)(M4),
    inner sep=4pt
] (gapbox) {};

\node[
    font=\scriptsize\itshape,
    text=gapred,
    below=2pt of gapbox.south,
    anchor=north
] {ad hoc / underdeveloped};

\draw[
    -{Stealth[length=4pt]},
    thick,
    black!40,
    shorten >=2pt,
    shorten <=2pt
] ($(P1.south west)+(-0.3,0.45)$) -- ($(P1.south west)+(-0.3,-0.45)$)
    node[midway, left, font=\tiny, align=right, text=black!60] {current\\focus};

\end{tikzpicture}  
\caption{\textbf{The robustness gap in frontier AI governance.} Risk of harm accumulates 
downstream (top row). Opportunities for prevention concentrate upstream. Broader mitigation capacity 
remains underdeveloped. However, this determines society's capacity to adopt and learn.}
\label{fig:coordination-gap}
\end{figure}
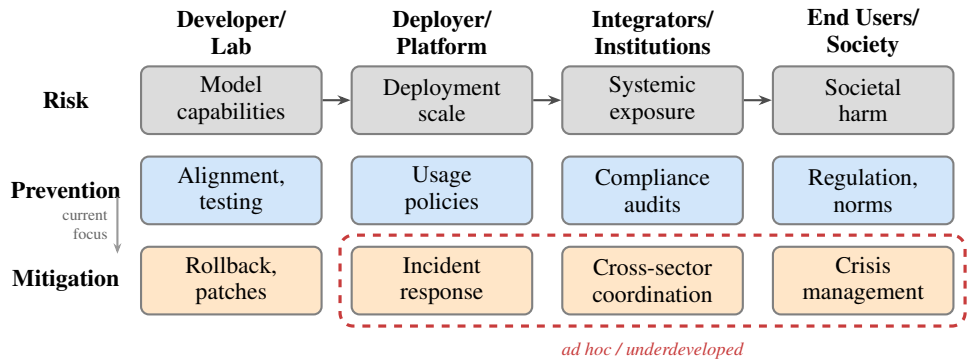

As Figure \ref{fig:coordination-gap} illustrates, existing AI governance efforts are heavily concentrated on prevention, with substantial work on alignment and testing \cite{amodei2016concrete}, usage constraints \cite{brundage2018malicious}, compliance auditing \cite{brundage2026frontier}, and regulatory norms \cite{EUAIAct2024,CaliforniaSB53}. By contrast, mitigation and preparedness remain far less developed. Although monitoring and incident reporting have begun to receive attention \cite{OECD2025aiincidents}, policy proposals addressing coordination and crisis management are sparse, leaving coordination capacity the most underexplored element of the governance landscape.

Robust Decision Making (RDM) offers an alternative framework. Rather than relying on contested forecasts, it stress-tests candidate strategies across hundreds of plausible futures to reveal vulnerabilities and adaptive levers. This shifts attention from prediction to preparation, and from ``what will happen?'' to ``what actions hold across many futures?'' and ``what tradeoffs are we willing to make?'' \cite{randRDMtool,lempert2013making}. Neglecting such robustness thinking in frontier AI governance means that Responsible Scaling Policies (RSPs) remain strong at local gating---controlling deployment decisions within a single organization---but weak at global adaptability. Without cross-domain triggers, external signposts, and pre-committed adaptation pathways, policies risk either paralysis or ad hoc responses once crises unfold.

\subsection{Tentative steps towards robust AI governance}

Current FASPs make effective use of adaptive triggers. Anthropic's AI Safety Level (ASL) upgrades, OpenAI's High/Critical risk gates, and DeepMind's Critical Capability Level (CCL) alerts provide clear internal triggers for escalating safety requirements \cite{openai_prepframework_v2_2025,anthropic2023rsp,meta_frontier_ai_2025}. Each lab has implemented some form of scalable or early-warning evaluation, along with post-deployment monitoring and red-team feedback loops. The policies contain firm pre-commitments to halt or withhold deployment absent required safeguards, with named decision bodies and defined documentation requirements such as capability reports and safety case assessments.

However, these mechanisms remain internally focused and siloed by hazard category. Consider the following cross-domain scenarios: cybersecurity disruptions that cascade into compromised biosafety protocols in research laboratories; AI agents deployed at scale to commit or assist in criminal activity; a model that successfully exfiltrates itself but fails to recursively self-improve, leaving the system in an ambiguous risk state; or significant psychological harms emerging in vulnerable subpopulations from sustained interaction with frontier models. Current frameworks lack explicit triggers and response protocols for such scenarios.

Recent proposals by OpenAI and Anthropic have begun to incorporate transparency and coordination as important criteria \cite{openai_scvd_2025, anthropic_transparency_2025}. The primary motivation appears to be ecosystem security, particularly regarding cybersecurity vulnerabilities. Anthropic proposes transparency requirements for frontier developers' safety policies and notes that required safeguards can be lowered if another actor passes a capability threshold without them. While this represents precisely the kind of robust, conditional, and transparent mechanism that is argued as necessary, it constitutes a relaxation of commitments rather than improving mitigation.

These developments acknowledge both the need for transparency and adaptive triggers, yet fall short of enabling coordination on optimal responses. In a crisis scenario, critical conversations will either be held for the first time or conducted without knowledge of other relevant actors' reasoning processes---precisely when coordination capacity matters most.

\section{Introducing robustness thinking to FASP and other stakeholders}

Integrating robustness into frontier AI governance requires three foundational shifts. First, reactions, tradeoffs, and reasoning in response to triggers must be made transparent---not merely the triggers themselves, but the decision logic that converts signals into actions. Second, triggers must be partially externalized, moving beyond purely technical metrics to incorporate consequentialist and external measures of societal impact. Third, this integration should build on existing institutional capacity: some government bodies already deploy robustness metrics in adjacent domains \cite{mangin2025robust}, providing templates for adaptation.

A growing body of work makes clear that evaluation alone is insufficient for managing high-stakes risk. \citet{lukovsiute2025llm} show that even sophisticated capability evaluations fail to capture real-world risk dynamics, particularly under adversarial or rapidly shifting operational conditions. This limitation is not merely technical: it reflects a broader structural challenge in which uncertainty about failure modes and their thresholds undermines coordinated precaution. Experimental evidence from climate threshold-public-goods games illustrates this dynamic starkly. When the location or consequences of a catastrophic threshold are uncertain, cooperative investment collapses--even among actors who prefer the cooperative outcome and understand the stakes \cite{barrett2014sensitivity}. The mechanism is simple: uncertainty about others’ responses lowers each actor’s expectation that coordinated action will materialize, producing individually rational but systemically disastrous underinvestment. Frontier-AI safety faces the same structural vulnerability. As long as actors hold divergent beliefs about thresholds of unacceptable risk and lack visibility into others’ intended responses, even well-designed preventive policies exhibit a robustness gap.\\

\subsection{Learning from other domains}

Risk management in other disciplines offers valuable lessons, though rarely at the scope and complexity of transformative AI \cite{stetler2025reinsuring}. The most relevant precedents come from nuclear safety, pandemic preparedness, and cybersecurity---domains that share characteristics of catastrophic tail risk, socio-technical complexity, and the need for coordinated response across organizational boundaries. These domains have converged on layered safeguards, rapid response capacity, early warning systems, pre-decided decision frameworks \cite{gai2010contagion}, and pre-positioned resources. Interdependence between critical infrastructure and their failure modes is well documented \cite{prier2023interdependence,sonesson2021governance,alcaraz2015critical}, and similar thinking will be necessary for AI crisis preparedness. Each domain has also developed a specific coordination device: Sector Risk Management Agencies (SRMAs) designate sector leads that convene owners and operators through standing information-sharing channels \cite{CISA_SRMA,yusta2011methodologies}; the International Health Regulations (IHR) impose mandatory, time-bounded signal escalation that creates common knowledge under uncertainty \cite{gostin2016international}; and the IAEA provides standardized incident classification vocabularies enabling rapid cross-actor interpretation without renegotiating meanings mid-crisis \cite{IAEA_NEA_INES_2001}.

Across these regimes, the shared institutional function is not ``perfect prevention'' but rather a set of coordination primitives: precommitment to notify and coordinate, interoperable procedures, and standing venues for joint situational awareness and response alignment \cite{perry2003preparedness}. In each case, coordination capacity is treated as a first-class governance objective---responsibilities for convening, escalation, and cross-actor alignment are assigned ex ante rather than improvised during crises \cite{christensen2016comparing}. What appears optimal from an individual actor's perspective may produce dramatically worse outcomes at the system level; individual organizations optimizing their own response plans in isolation may duplicate efforts inefficiently, create incompatible protocols, or leave critical gaps uncovered \cite{gai2010contagion}.

Advanced AI governance faces structural challenges that make these coordination problems more acute. First, the broad applicability of general-purpose technologies---particularly agentic systems---generates fuzzy domain boundaries, ever-diversifying taxonomies, and arbitrary thresholds that frustrate the standardization these regimes rely upon. Unlike nuclear incidents or disease outbreaks, AI-related harms often lack clear categorical boundaries: cybersecurity disruptions may cascade into compromised biosafety protocols; psychological harms may emerge gradually in vulnerable subpopulations; models may partially exfiltrate without achieving recursive self-improvement, leaving systems in ambiguous risk states. Second, increasing potential severity combined with shorter response times makes institutional learning-by-doing---the mechanism through which previous regimes developed their coordination capacity---undesirable or infeasible. Third, unlike established regimes that built trust and interoperability over decades of practice, AI governance must develop coordination capacity ahead of time, before the focusing events that typically catalyze institutional development. This means coordination mechanisms cannot wait for consensus on threat vectors or accumulated incident data; they must function under conditions of deep uncertainty about failure modes themselves \cite{somani2025_strengthening_rra3847-1}.

Transparency, rather than legal commitment, may be sufficient to substantially increase systemic coordination capacity. However, many interventions are costly, and private entities are understandably cautious about committing capital to contingent responses. This makes it valuable to first investigate the option space of possible interventions through structured scenario-response explorations, which can reveal synergies and complementarities in coordination capacity, or identify responses by one actor that are necessary to enable effective responses by others.

\subsection{A tentative policy proposal}

We propose a Scenario Response Registry (SRR) as an institutional mechanism to operationalize robustness thinking.\footnote{The SRR builds on the concept of risk registers, established in industrial risk management, which catalog known hazards alongside likelihood assessments, impact ratings, and designated responses \cite{aven2016risk}. The key difference is that the SRR requires prospective filing based on foresight about failure modes that have not yet materialized.} The core function is to make coordination capacity visible, comparable, and stress-testable before crises materialize. By requiring actors to articulate their intended responses to specified scenarios, the SRR enables red-teaming of governance frameworks, evaluation of policy under technical constraints, modelling of adversarial and cascading risk scenarios, and drafting of governance proposals anchored in real model behaviour. Where current approaches leave coordination to be improvised under crisis conditions, the SRR shifts learning and alignment upstream.

A public authority would maintain the registry, curating a library of salient technical, socio-economic, and security scenarios with specified potential harms. All relevant actors---frontier labs, cloud providers, critical infrastructure operators, and government agencies---would file ex ante ``if-then'' plans detailing thresholds that would trigger action, the specific actions to be taken, and resource commitments allocated to those responses. Submissions would be standardized, time-stamped, and comparable, creating common knowledge and curbing cheap talk---the costless announcement of commitments unlikely to be honored. The registry would review filings to identify gaps, flag overlaps, and suggest improvements. Access structures may be asymmetric: the scenario library could be broadly accessible while specific response filings and harmonization analyses remain restricted to relevant actors and the coordinating authority.

The SRR's incentive architecture would link disclosure quality to tangible consequences. Quality-weighted plans could determine regulatory forbearance, access to public compute resources, or procurement eligibility. Skin-in-the-game mechanisms---such as bonds or insurance instruments---would be forfeited upon failure to act when declared triggers are met. Dynamic scoring would update robustness assessments that shape audit intensity and capital requirements, creating ongoing incentives for plan quality and execution capacity. An independent technical panel with rotating membership would set the scenario library and auditing rubric, limiting regulatory capture while maintaining technical credibility.

\begin{figure}[H]
\centering
\begin{tikzpicture}[
    node distance=0.8cm and 0.5cm,
    stagebox/.style={
        rectangle,
        rounded corners=4pt,
        minimum width=2.8cm,
        minimum height=3.6cm,
        align=center,
        draw=black!60,
        thick
    },
    outcomebox/.style={
        rectangle,
        rounded corners=6pt,
        minimum width=3.2cm,
        minimum height=2.4cm,
        align=center,
        draw=black!70,
        very thick
    },
    stagetitle/.style={
        font=\footnotesize\bfseries,
        align=center
    },
    stagecontent/.style={
        font=\scriptsize,
        align=left,
        text width=2.4cm
    },
    mainarrow/.style={
        -{Stealth[length=6pt]},
        thick,
        arrowcol
    },
    feedbackarrow/.style={
        -{Stealth[length=5pt]},
        thick,
        feedbackcol,
        dashed
    },
    contribarrow/.style={
        -{Stealth[length=5pt]},
        thick,
        black!50
    },
    srrbox/.style={
        rectangle,
        rounded corners=8pt,
        draw=red,
        densely dotted,
        very thick,
        inner sep=0.4cm
    },
    inputbox/.style={
        rectangle,
        rounded corners=2pt,
        minimum width=1.8cm,
        minimum height=0.5cm,
        align=center,
        draw=black!40,
        fill=black!5,
        font=\tiny
    }
]
\node[stagebox, fill=scenariobg] (S1) at (0, 0) {};
\node[stagetitle, above=0.35cm of S1.center, anchor=south] (T1) {Scenario\\Library};
\node[stagecontent, below=0.05cm of T1.south, anchor=north] {%
    \textbullet~Technical\\
    \textbullet~Socio-economic\\
    \textbullet~Security\\
    \textbullet~Cross-domain%
};
\node[stagebox, fill=filingbg, right=1.2cm of S1] (S2) {};
\node[stagetitle, above=0.35cm of S2.center, anchor=south] (T2) {Actor\\Filings};
\node[stagecontent, below=0.05cm of T2.south, anchor=north] {%
    \textbf{If:} trigger met\\
    \textbf{Then:} action\\
    \textbf{With:} resources\\[2pt]
    {\tiny Labs, cloud, gov.,}\\
    {\tiny infrastructure}%
};
\node[stagebox, fill=srrbg, right=1.2cm of S2] (S3) {};
\node[stagetitle, above=0.35cm of S3.center, anchor=south] (T3) {Harmonization};
\node[stagecontent, below=0.05cm of T3.south, anchor=north] {%
    \textbullet~Identify gaps\\
    \textbullet~Flag overlaps\\
    \textbullet~Score plans\\
    \textbullet~Stress-test%
};
\node[srrbox, fit=(S1)(S2)(S3), inner ysep=0.6cm, label={[font=\footnotesize\bfseries, text=red]above:SRR}] (SRRbox) {};
\draw[mainarrow] (S1.east) -- (S2.west);
\draw[mainarrow] (S2.east) -- (S3.west);
\node[outcomebox, fill=crisisbg, below=1.4cm of S2] (CR) {};
\node[stagetitle, above=0.15cm of CR.center, anchor=south] {Crisis\\Preparedness};
\node[font=\scriptsize, below=0.0cm of CR.center, anchor=north, align=center, text width=2.8cm] {%
    {\tiny (improved outcome)}%
};
\node[inputbox, right=1.2cm of CR] (input1) {Monitoring capacity};
\node[inputbox, right=1.2cm of CR, yshift=-0.6cm] (input2) {Response capability};
\coordinate (right3) at ($(S3.east) + (0.6, 0)$);
\draw[contribarrow, rounded corners=8pt] 
    (S3.east) -- (right3) |- ([yshift=0.6cm]CR.east);

\node[font=\tiny, align=center, anchor=west]
    at ([yshift=0.9cm]input1.west)
    {Coordination Capacity};
\draw[contribarrow] (input1.west) -- (CR.east);
\draw[contribarrow] (input2.west) -- ([yshift=-0.6cm]CR.east);
\coordinate (left1) at ($(S1.west) + (-0.6, 0)$);
\draw[feedbackarrow, rounded corners=8pt] 
    (CR.west) -| (left1) -- (S1.west);
\node[font=\scriptsize\itshape, text=feedbackcol, align=center, rotate=90, anchor=south] 
    at ($(-1.5, -2.95)$) 
    {feedback};
\node[font=\tiny, text=black!50, above=0.15cm of S1.north] {1};
\node[font=\tiny, text=black!50, above=0.15cm of S2.north] {2};
\node[font=\tiny, text=black!50, above=0.15cm of S3.north] {3};
\end{tikzpicture}
\caption{\textbf{The Scenario Response Registry (SRR) as a coordination mechanism.} Scenarios are curated by an independent panel (1). Relevant actors file standardized if--then response plans (2). The SRR harmonizes filings to identify gaps and coordination opportunities (3). Together with complementary efforts—incident monitoring and preparedness frameworks—the SRR contributes to improved crisis readiness. Feedback from drills and real incidents updates the scenario library.}
\label{fig:srr-process}
\end{figure}
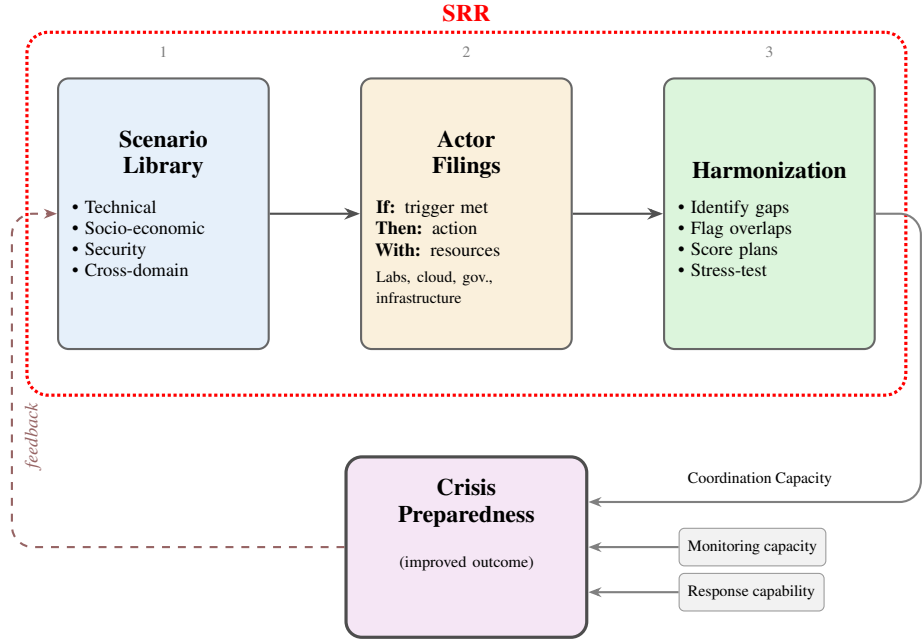

\section{Challenges and conditions for integration}

The SRR does not emerge in a vacuum. Complementary efforts in the broader ecosystem are addressing other dimensions of the preparedness deficit. Various incident databases document realized harms or near-misses, providing empirical grounding for scenario identification \cite{mcgregor2021preventing,mitre_attack}. Recent work develops operational frameworks for emergency response and loss of control \cite{somani2025_strengthening_rra3847-1, stix2025loss}. These contributions advance incident learning and response planning—but neither addresses how actors coordinate credibly to each other's responses before crises materialize. The SRR targets this coordination gap directly, complementing calls for greater systematicity in frontier AI risk management \cite{ziosi2025safety}.

\subsection{Why robustness is undersupplied}

The underinvestment in robustness thinking in frontier AI governance stems from a fundamental misalignment between private and social returns \cite{friedman2025shared}. Private returns favor speed and scope---first-mover advantages, market share, and strategic positioning---while social returns hinge on resilience and the capacity to absorb shocks. Firms therefore systematically underinvest in robustness \cite{anderljung2023frontier}. Strategic ambiguity compounds this problem by preserving bargaining power in negotiations with regulators and competitors. Codified triggers that expose tradeoffs and curtail managerial discretion are thus resisted, as they reduce flexibility that may prove valuable under competitive pressure \cite{selivanovskikh2025strategic}.

Translating abstract commitments into operational thresholds, triggers, and rules requires substantial upfront investment in scenario mapping and the identification of credible signposts \cite{marchau2019decision}. Miscalibration carries significant costs: false positives erode credibility and generate pressure to relax standards, while false negatives allow harm to accumulate undetected \cite{sawada2022impact, ec2023_precautionary_principle}. This risk of calibration failure reinforces organizational preferences for implicit pathways that preserve interpretive flexibility.

Maintaining live indicators demands sustained resources, disciplined data collection, and specialized analytic capacity. Without continuous investment, these systems decay into periodic check-ins that provide neither early warning nor actionable intelligence. This degradation further reduces the perceived payoff to robustness investments, creating a negative feedback loop.

Robustness exhibits the classic structure of a public good: benefits are diffuse across many stakeholders, while the costs are concentrated on those who must slow development, narrow deployment scope, or invest in standby capacity \cite{hardin2015collective}. This predictably generates free-riding and holdout problems, and measures that impose real constraints face resistance unless governance rules are perceived as fair, technically justified, and applied consistently across competitors. In a first-best setting, these incentive failures require strong governance institutions capable of verifying commitments, enforcing reciprocity, and resisting capture by either regulated entities or bureaucratic interests \cite{abbott2015international}. In their absence, however, coordination mechanisms should not be evaluated by whether they offer guarantees, but by whether they measurably improve coordination relative to the status quo. Absent such institutions, coordination-enhancing arrangements that increase transparency, align expectations, and support preparedness may not fully resolve the commons problem, but they can still dominate uncoordinated responses and reduce the risk that robustness failures propagate unchecked.\

\subsection{Towards improving coordination}

Several structural constraints limit the feasibility of global pre-coordination in frontier AI governance. Geopolitical dynamics and strategic rivalry reduce willingness to share assessments, thresholds, or vulnerabilities across jurisdictions \cite{kydd2005trust}. Competitive pressures create incentives for both states and firms to conceal strategically relevant capabilities, undermining the credibility of ex ante commitments \cite{lieber2017new}. Because sensitive technical information is inherently dual-use, any viable coordination regime must incorporate tiered access, controlled-sharing infrastructures, or delayed-release protocols \cite{national2004biotechnology, national2007science}.

These constraints suggest that global coordination---requiring convergence on shared threat models, uniform standards, and binding enforcement---is unrealistic under current conditions of strategic mistrust, asymmetric information, and divergent regulatory priorities. However, polycentric governance offers a viable alternative \cite{ostrom2010beyond, jones2020political}. Rather than awaiting comprehensive international agreement, coordination can begin locally through domestic SRRs, bilateral agreements such as US-UK arrangements or lab consortia, or sector-specific pilots.

A common objection holds that local or partial approaches risk irrelevance without broader participation. Yet durable coordination regimes often begin locally and expand through overlapping coalitions as trust accumulates \cite{keohane2011regime}. The key design principle is interoperability over uniformity: compatible procedures that can later integrate as mutual confidence builds \cite{chin2026interoperability}. Interoperability represents a lower bar than full harmonization and is therefore more robust under political uncertainty \cite{abbott2021governance}. Concrete bootstrapping pathways---whether anchored in lab consortia, cloud infrastructure providers, or bilateral regulatory arrangements---warrant dedicated analysis beyond the scope of this paper, though the interoperability principles would apply across configurations.

Any viable policy solution must function under realistic failure modes: partial participation, strategic under-specification, geopolitical fragmentation, and crisis-time friction. Critically, even partial coordination delivers value by reducing uncertainty about others' responses among willing actors \cite{cooper1992communication}. Tiered access regimes and controlled-sharing infrastructures can reconcile transparency with security constraints; such mechanisms are already deployed for hardware access and can apply similarly to information crucial for coordination \cite{RAND_Understanding}. 

A final objection notes that without enforcement, no mechanism can guarantee compliance. This is true but misses the point: imperfect coordination mechanisms implemented sooner dominate perfect mechanisms implemented later, particularly when institutional learning is required. The experience of the International Health Regulations illustrates this dynamic---even with established frameworks, implementation capacity lags behind formal commitments, arguing for starting now to build institutional muscle \cite{gostin2016international}. For a minimum viable SRR to be impactful requires only a curated scenario set, willing filers, a reviewing body, and regular tabletop exercises. These modest requirements lower the barrier to entry while establishing the infrastructure for more robust coordination as the ecosystem matures. 

The SRR itself is subject to failure modes that could undermine its coordination function. Scenario libraries may be misspecified---either too narrow to capture emergent cross-domain risks or too expansive to enable meaningful preparation. Harmonization processes could generate false confidence if actors treat filed plans as sufficient preparation rather than inputs to ongoing learning. Performative compliance remains a risk: organizations may file detailed responses they lack capacity or intent to execute. Finally, the scenario-setting process is vulnerable to capture, whether by regulated entities seeking to narrow the scope of plausible threats or by bureaucratic interests seeking to expand jurisdiction. These failure modes argue for iterative design with built-in review mechanisms rather than a fixed institutional architecture.

\section{Conclusion}
Frontier AI Safety Policies are built on the assumption that prevention will hold. But these policies also implicitly assume that when prevention fails, the broader ecosystem will know what to do, and at present it does not. There is no shared vocabulary for cross-domain failures, no agreed escalation path, and no common knowledge of how major actors would behave once their internal triggers fire. This is the coordination gap.
Closing it does not require global agreement, which is unlikely to arrive in time. It requires that the actors who already make these decisions, labs, infrastructure operators, and regulators, expose their if-then reasoning to one another ex ante. Such note-exchange is the cheapest form of coordination capacity available, and the form most likely to scale with the pace of capability progress, deployment, and incidents. Polycentric beginnings, domestic, bilateral, and sector-specific, are not a compromise on global ambitions; they are how durable regimes typically start.
The case for this work is not that it replaces prevention. It is that the institutions capable of learning at the speed of frontier AI do not yet exist, and waiting for the focusing events that would otherwise produce them is not a coordination strategy that anyone should choose. Building those institutions now, while the cost of doing so is a cost that can be chosen rather than imposed, is the agenda this paper opens.

\begin{ack}
The author is especially grateful to Jan Pieter Snoeij (Simon Institute for Longterm Governance) for sustained and substantive discussions that shaped both the framing and the central arguments of this paper. The author also thanks the Centre for the Governance of AI and co-fellows for valuable input and discussions. The author declares no competing interests.
\end{ack}

\bibliographystyle{plainnat}   
\bibliography{refs}

\end{document}